\documentclass[preprint]{ptephy}

\preprintnumber{KYUSHU-HET-147}

\usepackage{amssymb}
\usepackage{amsthm}
\usepackage{amsmath}
\usepackage{booktabs}
\DeclareMathOperator{\tr}{tr}
\DeclareMathOperator{\Tr}{Tr}

\newcommand{\Slash}[1]{{\ooalign{\hfil/\hfil\crcr$#1$}}}
\numberwithin{equation}{section}

\usepackage{url}
\let\Gamma\varGamma

\begin{document}

\title{Universal formula for the energy--momentum tensor via a flow equation in
the Gross--Neveu model}

\author{%
\name{\fname{Hiroshi} \surname{Suzuki}}{1,\ast}
}

\address{%
\affil{1}{Department of Physics, Kyushu University, 6-10-1 Hakozaki,
Higashi-ku, Fukuoka, 812-8581, Japan}
\email{hsuzuki@phys.kyushu-u.ac.jp}
}

\begin{abstract}
For the fermion field in the two-dimensional Gross--Neveu model, we introduce a
flow equation that allows a simple $1/N$ expansion. By employing the $1/N$
expansion, we examine the validity of a universal formula for the
energy--momentum tensor which is based on the small flow-time expansion. We
confirm that the formula reproduces a correct normalization and the
conservation law of the energy--momentum tensor by computing the translation
Ward--Takahashi relation in the leading non-trivial order in the
$1/N$ expansion. Also, we confirm that the expectation value at finite
temperature correctly reproduces thermodynamic quantities. These observations
support the validity of a similar construction of the energy--momentum tensor
via the gradient/Wilson flow in lattice gauge theory.
\end{abstract}
\subjectindex{B31, B32, B34, B38}
\maketitle

\section{Introduction}
\label{sec:1}
It has been well recognized~\cite{Caracciolo:1988hc,Caracciolo:1989pt} that the
construction of the energy--momentum tensor---the Noether current associated
with the translational invariance---is quite involved in lattice field theory.
This is because lattice regularization explicitly breaks the translational
invariance and the energy--momentum tensor is a composite operator containing
local products of field variables. Because of radiative corrections, a naive
discretization as it stands cannot reproduce a correct normalization and the
conservation law of the energy--momentum tensor in the continuum limit.
Recently, a completely new approach to this problem, on the basis of the
gradient/Wilson flow~\cite{Luscher:2010iy,Luscher:2013cpa,Luscher:2013vga} and
the small flow-time expansion~\cite{Luscher:2011bx}, has been
proposed~\cite{Suzuki:2013gza,DelDebbio:2013zaa,Makino:2014taa} in the
context of lattice gauge theory. In that approach, especially that
in~Refs.~\cite{Suzuki:2013gza,Makino:2014taa}, one constructs a ``universal
formula'' for the energy--momentum tensor using a perturbative solution of the
gradient flow. This construction relies on the UV finiteness of the gradient
flow in gauge theory~\cite{Luscher:2011bx,Luscher:2013cpa} such that
renormalization of any composite operator of flowed fields is very simple. The
universal formula is supposed to provide a regularization-independent
expression for the energy--momentum tensor and thus is expected to be usable
even with lattice regularization.

The above approach is based on natural assumptions such as the existence of the
energy--momentum tensor and the renormalizability of the gradient flow in the
non-perturbative level. Also, the formula in~Ref.~\cite{Suzuki:2013gza} has
been numerically tested for quenched QCD at finite
temperature~\cite{Asakawa:2013laa,Kitazawa:2014uxa}. However, it still remains
important to investigate the validity of the approach in various possible ways.
In particular, it is of great interest whether and how the universal formulas
in~Refs.~\cite{Suzuki:2013gza,Makino:2014taa}, which are constructed by using
perturbation theory, can capture non-perturbative low-energy physics or not.

As shown in~Ref.~\cite{Makino:2014sta}, the gradient flow in the
two-dimensional $O(N)$ non-linear sigma model~\cite{Kikuchi:2014rla} possesses
a UV finiteness quite similar to that of four-dimensional gauge theory. By
utilizing this UV finiteness, one can imitate the above construction of the
universal formula~\cite{Makino:2014sta}. For the two-dimensional $O(N)$
non-linear sigma model, the $1/N$ expansion is available and, to some extent,
the gradient flow can also be solved in the large~$N$
limit~\cite{Makino:2014cxa,Aoki:2014dxa}. In~Ref.~\cite{Makino:2014cxa}, using
this non-perturbative solution, the universal formula for the energy--momentum
tensor has been analytically tested by computing the expectation value at
finite temperature. The expectation value correctly reproduces thermodynamic
quantities obtained by the conventional $1/N$ expansion. This study demonstrates
that the universal formula reproduces a correct normalization at least for
those quantities.

Another interesting issue is whether the conservation law (and more general
Ward--Takahashi relations associated with the translational invariance) is
correctly reproduced by the universal formula. This analysis for the
two-dimensional $O(N)$ non-linear sigma model has not been carried out, because
in~Ref.~\cite{Makino:2014cxa} the gradient flow was solved only in the leading
order in the $1/N$ expansion with which any correlation function is factorized
into one-point functions.\footnote{It might be possible to use the large $N$
solution given in~Ref.~\cite{Aoki:2014dxa} to investigate this issue.}

In the present paper, with the above motivations, we consider a similar
universal formula for the energy--momentum tensor in the two-dimensional
Gross--Neveu model~\cite{Gross:1974jv}. The point is that, in this non-gauge,
unconstrained system, one can introduce a very simple flow equation that does
not contain any interaction. Although this is not the gradient flow in the
sense that the flow is defined with respect to the equation of motion of the
original system, such a choice is perfectly legitimate from the perspective of
the UV finiteness of the flow. Similar simplification has also been adopted for
the flow of the fermion field in gauge theory~\cite{Luscher:2013cpa}. Because
of this simplification in the flow equation, the conventional $1/N$
expansion~\cite{Gross:1974jv,Coleman:1980nk} directly provides the solution of
the flowed fields. We can then readily examine, in the leading non-trivial
order in the $1/N$ expansion, if the universal formula correctly reproduces the
translation Ward--Takahashi relation 

This paper is organized as follows. In Sect.~\ref{sec:2}, we introduce a flow
equation for the fermion field in the Gross--Neveu model. In~Sect.~\ref{sec:3},
along the line of reasoning
in~Refs.~\cite{Suzuki:2013gza,Makino:2014taa,Makino:2014sta}, we construct a
universal formula for the energy--momentum tensor in the present system. This
construction itself is based on one-loop matching with the expression with
dimensional regularization. In~Sect.~\ref{sec:4}, we recapitulate the
conventional $1/N$ expansion of the present system. Section~\ref{sec:5} is the
main part of the paper and, in the leading non-trivial order in the $1/N$
expansion, we examine if the universal formula correctly reproduces (some
particular cases of) the translation Ward--Takahashi relation. Here, we observe
that the universal formula precisely reproduces expected relations with the
presence of the non-perturbative mass gap, although the construction of the
universal formula itself uses one-loop perturbation theory. As another support
for the universal formula, in~Sect.~\ref{sec:6}, we compute the expectation
value of the energy--momentum tensor defined by the universal formula at finite
temperature as~Ref.~\cite{Makino:2014cxa}. It reproduces the correct results.
The last section is devoted to conclusions.

\section{Flow equation in the Gross--Neveu model}
\label{sec:2}
The Euclidean action of the Gross--Neveu model~\cite{Gross:1974jv} is given by
\begin{equation}
   S=\int\mathrm{d}^Dx\,\left\{
   \Bar{\psi}^i(x)\Slash{\partial}\psi^i(x)
   -\frac{\lambda_0}{2N}\left[\Bar{\psi}^i(x)\psi^i(x)\right]^2\right\},
\label{eq:(2.1)}
\end{equation}
where $D=2$ for our target theory and the fermion field has $N$ components
($i=1$, $2$, \dots, $N$).\footnote{The summation over repeated ``flavor''
indices~$i$, $j$, \dots, and Lorentz indices~$\mu$, $\nu$, \dots, is always
understood in this paper.} In this system, we introduce a flow equation. That
is, we introduce a fictitious time~$t$ and suppose that the fermion field
evolves according to
\begin{align}
   \partial_t\chi(t,x)=\partial_\mu\partial_\mu\chi(x),\qquad
   \chi(t=0,x)=\psi(x),
\label{eq:(2.2)}
\\
   \partial_t\Bar{\chi}(t,x)=\partial_\mu\partial_\mu\Bar{\chi}(x),\qquad
   \Bar{\chi}(t=0,x)=\Bar{\psi}(x),
\label{eq:(2.3)}
\end{align}
where the initial value for the evolution is given by the original fermion
field which is the subject of the functional integral (with the distribution
defined by~Eq.~\eqref{eq:(2.1)}). Note that Eqs.~\eqref{eq:(2.2)}
and~\eqref{eq:(2.3)} are very simple; the right-hand sides are defined by the
free Laplacian without any interaction. This is possible for the present
non-gauge, unconstrained system. Although the above flow is not the gradient
flow in the sense that the flow is defined by the equation of motion for the
original action~\eqref{eq:(2.1)}, such a choice is completely legitimate as far
as a UV finiteness of the flow is concerned---see the following discussions.

Since flow equations~\eqref{eq:(2.2)} and~\eqref{eq:(2.3)} do not contain any
interaction, the flowed fermion field becomes a simple linear functional of the
fermion field at the zero flow time. That is,
\begin{align}
   \chi(t,x)=\int\mathrm{d}^Dy\,K_t(x-y)\psi(y),\qquad
   \Bar{\chi}(t,x)=\int\mathrm{d}^Dy\,K_t(x-y)\Bar{\psi}(y),   
\label{eq:(2.4)}
\end{align}
where\footnote{Throughout this paper, we use the abbreviation
\begin{equation}
   \int_p\equiv\int\frac{\mathrm{d}^Dp}{(2\pi)^D}.
\label{eq:(2.5)}
\end{equation}
}
\begin{equation}
   K_t(x)=\int_p\mathrm{e}^{ipx}\mathrm{e}^{-tp^2}
   =\frac{\mathrm{e}^{-x^2/4t}}{(4\pi t)^{D/2}}
\label{eq:(2.6)}
\end{equation}
is the heat kernel for the free Laplacian. Using Eq.~\eqref{eq:(2.4)},
correlation functions of the flowed fermion field can directly be obtained in
terms of correlation functions of the original fermion field. For example,
since the tree-level propagator of the original fermion field is given by
\begin{equation}
   \left\langle\psi^i(x)\Bar{\psi}^j(y)\right\rangle_0
   =\delta^{ij}\int_p\mathrm{e}^{ip(x-y)}\,
   \frac{1}{i\Slash{p}},
\label{eq:(2.7)}
\end{equation}
the tree-level propagator of the flowed field is
\begin{equation}
   \left\langle\chi^i(t,x)\Bar{\chi}^j(s,y)\right\rangle_0
   =\delta^{ij}\int_p\mathrm{e}^{ip(x-y)}\,
   \frac{\mathrm{e}^{-(t+s)p^2}}{i\Slash{p}}.
\label{eq:(2.8)}
\end{equation}
Also, the renormalization property of the unflowed fermion field is directly
inherited by the flowed fermion field. In particular, their wave function
renormalization constants are identical. This is quite different from the
flowed fermion field in gauge theory~\cite{Luscher:2013cpa} in which the wave
function renormalization constant for the flowed fermion field is independent
of that of the original fermion field, due to interaction in the flow equation.

\section{Universal formula for the energy--momentum tensor}
\label{sec:3}
In this section, following the idea
of~Refs.~\cite{Suzuki:2013gza,Makino:2014taa,Makino:2014sta}, we construct a
universal formula for the energy--momentum tensor in the Gross--Neveu
model~\eqref{eq:(2.1)} using the small flow-time
expansion~\cite{Luscher:2011bx}. We first assume dimensional regularization
with~$D=2-\epsilon$ and derive the explicit form of the energy--momentum
tensor. Since dimensional regularization preserves the translational
invariance, that energy--momentum tensor fulfills the Ward--Takahashi relation
associated with the translational invariance; this implies that the
energy--momentum tensor is correctly normalized and is conserved. However, the
energy--momentum tensor with dimensional regularization is useful only in
perturbation theory. Our universal formula below is intended to provide a
regularization-independent expression for the energy--momentum tensor. This
universal formula is thus also expected to be usable with lattice
regularization for example, with which non-perturbative calculations are
possible.

Assuming dimensional regularization, the energy--momentum tensor can be
obtained from the variation of the action
\begin{equation}
   \delta S=-\int\mathrm{d}^Dx\,\xi_\nu(x)\partial_\mu T_{\mu\nu}(x)
\label{eq:(3.1)}
\end{equation}
under the transformations
\begin{equation}
   \delta\psi(x)=\xi_\mu(x)\partial_\mu\psi(x),\qquad
   \delta\Bar{\psi}(x)=\xi_\mu(x)\partial_\mu\Bar{\psi}(x).
\label{eq:(3.2)}
\end{equation}
The explicit form is given by\footnote{Here, we have taken only the part of the
expression appearing in~Eq.~\eqref{eq:(3.1)} being symmetric
under~$\mu\leftrightarrow\nu$; the anti-symmetric part generates the Lorentz
transformation and is not explicitly considered in what follows.}
\begin{equation}
   T_{\mu\nu}(x)
   =\frac{1}{4}\Bar{\psi}^i(x)
   \left(\gamma_\mu\overleftrightarrow{\partial}_\nu
   +\gamma_\nu\overleftrightarrow{\partial}_\mu\right)
   \psi^i(x)
   -\delta_{\mu\nu}
   \left\{
   \Bar{\psi}^i(x)\frac{1}{2}\overleftrightarrow{\Slash{\partial}}\psi^i(x)
   -\frac{\lambda_0}{2N}\left[\Bar{\psi}^i(x)\psi^i(x)\right]^2
   \right\},
\label{eq:(3.3)}
\end{equation}
where $\overleftrightarrow{\partial}_\mu\equiv\partial_\mu
-\overleftarrow{\partial}_\mu$. This operator does not receive the
multiplicative renormalization, because of the translation Ward--Takahashi
relation
\begin{equation}
   \left\langle\mathcal{O}_{\text{ext}}
   \int_{\mathcal{D}}
   \mathrm{d}^Dx\,\partial_\mu
   T_{\mu\nu}(x)\,\mathcal{O}_{\text{int}}\right\rangle
   =-\left\langle\mathcal{O}_{\text{ext}}\,\partial_\nu\mathcal{O}_{\text{int}}
   \right\rangle,
\label{eq:(3.4)}
\end{equation}
where $\mathcal{D}$ is a bounded integration region, $\mathcal{O}_{\text{ext}}$
is an operator outside the region and $\mathcal{O}_{\text{int}}$ is an operator
inside the region. We define a renormalized energy--momentum tensor by
subtracting the (potentially UV-divergent) vacuum expectation value as
\begin{equation}
   \left\{T_{\mu\nu}\right\}_R(x)
   \equiv T_{\mu\nu}(x)-\left\langle T_{\mu\nu}(x)\right\rangle.
\label{eq:(3.5)}
\end{equation}

Now, to derive the universal formula, we express the composite
operator~\eqref{eq:(3.3)} in terms of the composite operator of the flowed
fermion field. This can be archived by the so-called small flow-time expansion
in~Ref.~\cite{Luscher:2011bx}. By a one-loop perturbative calculation similar to
that of~Refs.~\cite{Suzuki:2013gza,Makino:2014taa,Makino:2014sta}, we find
\begin{align}
   &\Bar{\chi}^i(t,x)
   \left(\gamma_\mu\overleftrightarrow{\partial}_\nu
   +\gamma_\nu\overleftrightarrow{\partial}_\mu\right)
   \chi^i(t,x)
   -\left\langle
   \Bar{\chi}^i(t,x)
   \left(\gamma_\mu\overleftrightarrow{\partial}_\nu
   +\gamma_\nu\overleftrightarrow{\partial}_\mu\right)
   \chi^i(t,x)
   \right\rangle
\notag
\\
   &=\Bar{\psi}^i(x)
   \left(\gamma_\mu\overleftrightarrow{\partial}_\nu
   +\gamma_\nu\overleftrightarrow{\partial}_\mu\right)
   \psi^i(x)
   -\frac{\lambda_0}{N}
   \frac{\lambda_0}{\pi}\left[\frac{2}{\epsilon}+\ln(8\pi t)+1\right]
   \delta_{\mu\nu}\left[\Bar{\psi}^i(x)\psi^i(x)\right]^2+O(t)
\label{eq:(3.6)}
\end{align}
and
\begin{align}
   &\left[\Bar{\chi}^i(t,x)\chi^i(t,x)\right]^2
   -\left\langle
   \left[\Bar{\chi}^i(t,x)\chi^i(t,x)\right]^2
   \right\rangle   
\notag
\\
   &=
   \left\{1-\frac{\lambda_0}{\pi}
   \left[\frac{2}{\epsilon}+\ln(8\pi t)\right]\right\}
   \left[\Bar{\psi}^i(x)\psi^i(x)\right]^2+O(t).
\label{eq:(3.7)}
\end{align}
In this and following one-loop computations, we retain only terms leading in
the large $N$ limit, because only leading terms are relevant in the analyses in
the following sections. From~Eq.~\eqref{eq:(3.6)}, we also have
\begin{align}
   &\Bar{\chi}^i(t,x)
   \overleftrightarrow{\Slash{\partial}}
   \chi^i(t,x)
   -\left\langle
   \Bar{\chi}^i(t,x)
   \overleftrightarrow{\Slash{\partial}}
   \chi^i(t,x)
   \right\rangle   
\notag
\\
   &=\Bar{\psi}^i(x)
   \overleftrightarrow{\Slash{\partial}}
   \psi^i(x)
   -\frac{\lambda_0}{N}
   \frac{\lambda_0}{\pi}\left[\frac{2}{\epsilon}+\ln(8\pi t)\right]
   \left[\Bar{\psi}^i(x)\psi^i(x)\right]^2+O(t).
\label{eq:(3.8)}
\end{align}

The relations~\eqref{eq:(3.6)}, \eqref{eq:(3.7)}, and~\eqref{eq:(3.8)} may be
inverted for composite operators of the unflowed fermion field. We then
substitute those expressions in~Eq.~\eqref{eq:(3.3)} to yield
\begin{align}
   \left\{T_{\mu\nu}\right\}_R(x)&=
   \frac{1}{4}\Bar{\chi}^i(t,x)
   \left(\gamma_\mu\overleftrightarrow{\partial}_\nu
   +\gamma_\nu\overleftrightarrow{\partial}_\mu\right)
   \chi^i(t,x)
   -\frac{1}{2}\delta_{\mu\nu}
   \Bar{\chi}^i(t,x)\overleftrightarrow{\Slash{\partial}}\chi^i(t,x)
\notag
\\
   &\qquad{}
   +\frac{\lambda_0}{2N}
   \left\{1+\frac{\lambda_0}{2\pi}
   \left[\frac{2}{\epsilon}+\ln(8\pi t)+1\right]\right\}
   \delta_{\mu\nu}\left[\Bar{\chi}^i(t,x)\chi^i(t,x)\right]^2-\text{VEV}
\notag
\\
   &\qquad\qquad+O(t),
\label{eq:(3.9)}
\end{align}
where $\text{VEV}$ denotes the vacuum expectation value of the composite
operator appearing in the right-hand side. In the one-loop order, the coupling
constant is renormalized in the minimal subtraction (MS) scheme as
\begin{equation}
   \lambda_0=\mu^\epsilon\lambda
   \left(1-\frac{\lambda}{\pi}\frac{1}{\epsilon}\right).
\label{eq:(3.10)}
\end{equation}
Then in terms of the renormalized coupling~$\lambda$, we have
\begin{align}
   \left\{T_{\mu\nu}\right\}_R(x)&=
   c_1(\lambda;\mu)\frac{1}{4}\Bar{\chi}^i(t,x)
   \left(\gamma_\mu\overleftrightarrow{\partial}_\nu
   +\gamma_\nu\overleftrightarrow{\partial}_\mu\right)
   \chi^i(t,x)
   -c_2(\lambda;\mu)\frac{1}{2}\delta_{\mu\nu}
   \Bar{\chi}^i(t,x)\overleftrightarrow{\Slash{\partial}}\chi^i(t,x)
\notag
\\
   &\qquad{}
   +c_3(\lambda;\mu)
   \delta_{\mu\nu}\left[\Bar{\chi}^i(t,x)\chi^i(t,x)\right]^2-\text{VEV}
\notag
\\
   &\qquad\qquad{}+O(t),
\label{eq:(3.11)}
\end{align}
where
\begin{align}
   c_1(\lambda;\mu)&=c_2(\lambda;\mu)=1+O(\lambda^2),
\label{eq:(3.12)}
\\
   c_3(\lambda;\mu)&=\frac{\lambda}{2N}
   \left\{1+\frac{\lambda}{2\pi}
   \left[\ln(8\pi\mu^2 t)+1\right]\right\}.
\label{eq:(3.13)}
\end{align}
As we have noted, in the present system, the flowed fermion field receives the
wave function renormalization common to the unflowed fermion field. Since the
fermion field does not receive the wave function renormalization to the
one-loop order in the present system,\footnote{This persists also in the
leading order $1/N$ expansion that is relevant to our analyses below. Thus,
in this paper, we do not need to consider the wave function renormalization of
the flowed fermion field. In gauge theory, on the other hand, renormalization
of the fermion field has to be taken into account; see
Ref.~\cite{Makino:2014taa}.} even \emph{composite operators\/} of the bare
flowed fermion field are UV finite without multiplicative renormalization; the
flow ensures this UV finiteness. Then Eqs.~\eqref{eq:(3.12)}
and~\eqref{eq:(3.13)} show that the right-hand side of~Eq.~\eqref{eq:(3.11)} is
UV finite. This should be so, because the energy--momentum tensor (after
subtracting the vacuum expectation value) in the left-hand side must be UV
finite.

Finally, we utilize a renormalization group argument. We apply the operation
\begin{equation}
   \left(\mu\frac{\partial}{\partial\mu}\right)_0
\label{eq:(3.14)}
\end{equation}
to both sides of~Eq.~\eqref{eq:(3.11)}, where the subscript~$0$ implies that
the bare quantities are kept fixed under the derivative. Since
the energy--momentum tensor~\eqref{eq:(3.3)} is entirely given by bare
quantities and the composite operators in the right-hand side
of~Eq.~\eqref{eq:(3.11)} are also bare, we infer that
$(\mu\partial/\partial\mu)_0c_i(\lambda;\mu)=0$ for~$i=1$, $2$, and~$3$. These
equations say that $c_i(\lambda;\mu)=c_i(\Bar{\lambda}(q);q)$ for
arbitrary~$q$, where $\Bar{\lambda}(q)$ is the running coupling in the MS
scheme with the renormalization scale~$q$. Since the renormalization scale~$q$
in~$c_i(\Bar{\lambda}(q);q)$ is arbitrary, we may take $q=1/\sqrt{8t}$ by using
the flow time~$t$. Then, since $\Bar{\lambda}(1/\sqrt{8t})\to0$ for~$t\to0$ by
the asymptotic freedom, the above perturbative computation is justified
for~$t\to0$. In this way, we arrive at
\begin{equation}
   \left\{T_{\mu\nu}\right\}_R(x)
   =\lim_{t\to0}\left[\Hat{T}_{\mu\nu}(t,x)
   -\left\langle\Hat{T}_{\mu\nu}(t,x)\right\rangle\right],
\label{eq:(3.15)}
\end{equation}
where
\begin{align}
   \Hat{T}_{\mu\nu}(t,x)&\equiv
   \frac{1}{4}\Bar{\chi}^i(t,x)
   \left(\gamma_\mu\overleftrightarrow{\partial}_\nu
   +\gamma_\nu\overleftrightarrow{\partial}_\mu\right)
   \chi^i(t,x)
   -\frac{1}{2}\delta_{\mu\nu}
   \Bar{\chi}^i(t,x)\overleftrightarrow{\Slash{\partial}}\chi^i(t,x)
\notag
\\
   &\qquad{}
   +\frac{\Bar{\lambda}(1/\sqrt{8t})}{2N}
   \left[1+\frac{\Bar{\lambda}(1/\sqrt{8t})}{2\pi}(\ln\pi+1)\right]
   \delta_{\mu\nu}\left[\Bar{\chi}^i(t,x)\chi^i(t,x)\right]^2.
\label{eq:(3.16)}
\end{align}
This is our universal formula for the energy--momentum tensor. This is
universal in the sense that it does not refer to any specific regularization;
the composite operator in the right-hand side is a renormalized quantity that
must be independent of regularization as far as the parameters are properly
renormalized.

We stress that our computation which led to~Eq.~\eqref{eq:(3.16)} is purely
one-loop. Although we retained only large $N$ leading terms in one-loop
coefficients, no non-perturbative $1/N$ expansion is invoked at this stage. In
particular, the fermion is treated as massless. We stress this point because
the intention of the present paper is to see how the formula~\eqref{eq:(3.16)}
that is obtained by one-loop perturbation theory can capture non-perturbative
physics. More specifically, we want to see if the idea that coefficients in the
universal formula can be determined by perturbation theory while low-energy
non-perturbative physics is contained in matrix elements of composite operators
works or not. This is the idea for the construction of the lattice
energy--momentum tensor
in~Refs.~\cite{Suzuki:2013gza,Makino:2014taa,Makino:2014sta}.

\section{$1/N$ expansion in the Gross--Neveu model}
\label{sec:4}
Now, for the analyses in subsequent sections, we briefly recapitulate the
well-known non-perturbative solution in the present system~\eqref{eq:(2.1)}, an
expansion in powers of~$1/N$~\cite{Gross:1974jv,Coleman:1980nk}.

For a systematic $1/N$ expansion, it is convenient to introduce an auxiliary
field~$\sigma(x)$ and rewrite the action~\eqref{eq:(2.1)} as
\begin{equation}
   S=\int\mathrm{d}^Dx\,\left[
   \Bar{\psi}^i(x)\Slash{\partial}\psi^i(x)
   +\sigma(x)\Bar{\psi}^i(x)\psi^i(x)
   +\frac{N}{2\lambda_0}\sigma(x)^2
   \right].
\label{eq:(4.1)}
\end{equation}
If we first integrate over the fermion field, the partition function becomes
\begin{equation}
   \mathcal{Z}=\int\left[\prod_x\mathrm{d}\sigma(x)\right]\,
   \exp\left\{
   -\frac{N}{2\lambda_0}\int\mathrm{d}^Dx\,\sigma(x)^2
   +N\Tr\ln\left[\Slash{\partial}+\sigma(x)\right]\right\}.
\label{eq:(4.2)}
\end{equation}
Since the exponent is proportional to~$N$ in this expression, in the leading
order of the $1/N$ expansion, the integral over the auxiliary field can be
approximated by the value at the saddle point. The saddle point is specified by
the stationary condition for the exponent, i.e., by the gap equation,
\begin{equation}
   \frac{1}{\lambda_0}\sigma=\tr\int_p\frac{1}{i\Slash{p}+\sigma}.
\label{eq:(4.3)}
\end{equation}
The momentum integration in the right-hand side requires regularization. If we
use dimensional regularization with~$D=2-\epsilon$, we have
\begin{equation}
   \frac{1}{\lambda_0}\sigma
   =\frac{1}{\pi}
   \left[\frac{1}{\epsilon}
   -\frac{1}{2}\ln\left(\frac{\mathrm{e}^\gamma\sigma^2}{4\pi}\right)\right]
   \sigma,
\label{eq:(4.4)}
\end{equation}
where $\gamma$ is the Euler constant. This tells us that, setting
\begin{equation}
   \lambda_0=\mu^\epsilon\lambda Z,
\label{eq:(4.5)}
\end{equation}
the renormalization factor is given by
\begin{equation}
   Z^{-1}=1+\frac{\lambda}{\pi}\frac{1}{\epsilon}
\label{eq:(4.6)}
\end{equation}
in the MS scheme. In terms of the renormalized coupling~$\lambda$, the saddle
point is expressed as
\begin{equation}
   \sigma^2=4\pi\mathrm{e}^{-\gamma}\Lambda^2,\qquad
   \Lambda\equiv\mu\mathrm{e}^{-\pi/\lambda}.
\label{eq:(4.7)}
\end{equation}
As Eq.~\eqref{eq:(4.1)} shows, this saddle point provides a non-perturbative
mass gap for the (originally massless) fermion. Corresponding
to~Eq.~\eqref{eq:(4.6)}, the beta function is given by
\begin{equation}
   \beta\equiv
   \left(\mu\frac{\partial}{\partial\mu}\right)_0\lambda
   =-\epsilon\lambda-\frac{\lambda^2}{\pi}
\label{eq:(4.8)}
\end{equation}
and thus the running coupling in the MS scheme is
\begin{equation}
   \Bar{\lambda}(q)=-\frac{2\pi}{\ln(\Lambda^2/q^2)}
   =-\frac{2\pi}{\ln[e^\gamma\sigma^2/(4\pi q^2)]}.
\label{eq:(4.9)}
\end{equation}

To obtain the next-to-leading order corrections in the $1/N$ expansion, we have
to consider the integration over the fluctuation around the saddle point
in~Eq.~\eqref{eq:(4.2)}. So we set
\begin{equation}
   \sigma(x)=\sigma+\delta\sigma(x).
\label{eq:(4.10)}
\end{equation}
The expansion of the exponent in~Eq.~\eqref{eq:(4.2)} is then
\begin{align}
   &-\frac{N}{2\lambda_0}\int\mathrm{d}^Dx\,\sigma(x)^2
   +N\Tr\ln\left[\Slash{\partial}+\sigma(x)\right]
\notag
\\
   &=-\frac{N}{2\lambda_0}\int\mathrm{d}^Dx\,\sigma^2
   +N\int\mathrm{d}^Dx\,\tr\int_p\ln(i\Slash{p}+\sigma)
\notag
\\
   &\qquad{}
   -\frac{N}{2\lambda_0}\int\mathrm{d}^Dx\,\delta\sigma(x)^2
\notag
\\
   &\qquad\qquad{}
   -\frac{N}{2}\int\mathrm{d}^Dx\int\mathrm{d}^Dy\,
   \delta\sigma(x)\delta\sigma(y)\int_p\mathrm{e}^{ip(x-y)}
   \tr\int_\ell\frac{1}{i\Slash{\ell}+\sigma}
   \frac{1}{i(\Slash{\ell}-\Slash{p})+\sigma}+O(\delta\sigma^3).
\label{eq:(4.11)}
\end{align}
There is no $O(\delta\sigma)$ term because $\sigma$ is the saddle point. After
the momentum integration and the parameter renormalization~\eqref{eq:(4.5)}, we
have
\begin{align}
   &-\frac{N}{2\lambda_0}\int\mathrm{d}^Dx\,\sigma(x)^2
   +N\Tr\ln\left[\Slash{\partial}+\sigma(x)\right]
\notag
\\
   &=\frac{N}{4\pi}\int\mathrm{d}^Dx\,\sigma^2
   -\frac{N}{4\pi}\int\mathrm{d}^Dx\int\mathrm{d}^Dy\,
   \delta\sigma(x)\delta\sigma(y)
   \int_p\mathrm{e}^{ip(x-y)}B(p^2,\sigma^2)
   +O(\delta\sigma^3),
\label{eq:(4.12)}
\end{align}
where~\cite{Gross:1974jv}
\begin{equation}
   B(p^2,\sigma^2)
   \equiv\sqrt{\frac{p^2+4\sigma^2}{p^2}}
   \ln\left(
   \frac{\sqrt{p^2+4\sigma^2}+\sqrt{p^2}}
   {\sqrt{p^2+4\sigma^2}-\sqrt{p^2}}\right).
\label{eq:(4.13)}
\end{equation}
From this, the propagator of the fluctuating field~$\delta\sigma(x)$ is given
by
\begin{equation}
   \left\langle\delta\sigma(x)\delta\sigma(y)\right\rangle
   =\frac{2\pi}{N}\int_p\mathrm{e}^{ip(x-y)}B(p^2,\sigma^2)^{-1}+O(1/N^2).
\label{eq:(4.14)}
\end{equation}

In the above computation, we may use lattice regularization as
well~\cite{Aoki:1983qi,Aoki:1985jj}. With the lattice spacing~$a$, one may
discretize the action~\eqref{eq:(4.1)} by replacing the Dirac
operator~$\Slash{\partial}$ by the Wilson Dirac operator for example,
\begin{equation}
   \frac{1}{2}
   \left[\gamma_\mu(\partial_\mu+\partial_\mu^*)
   -a\partial_\mu^*\partial_\mu\right]+m_0,
\label{eq:(4.15)}
\end{equation}
where $\partial_\mu$ and~$\partial_\mu^*$ are forward and backward difference
operators, respectively; $m_0$ is the bare mass parameter to be tuned to
restore the chiral symmetry explicitly broken by the Wilson term. Then setting
\begin{equation}
   \Tilde\sigma\equiv\sigma+m_0,
\label{eq:(4.16)}
\end{equation}
the gap equation with lattice regularization reads
\begin{align}
   \frac{1}{\lambda_0^{\text{LAT}}}(\Tilde{\sigma}-m_0)
   &=\tr\int_{\mathcal{B}}\frac{\mathrm{d}^2p}{(2\pi)^2}
   \frac{1}{i\Slash{\mathring{p}}+\frac{1}{2}a\Hat{p}^2+\Tilde{\sigma}}
\notag\\
   &=0.7698\,\frac{1}{a}
   -\frac{1}{2\pi}\left[\ln(a^2\Tilde{\sigma}^2)+1.11861\right]\Tilde{\sigma},
\label{eq:(4.17)}
\end{align}
where $\lambda_0^{\text{LAT}}$ is the bare coupling with lattice regularization,
$\mathcal{B}$ is the Brillouin zone
$\mathcal{B}\equiv\{p_\mu\mid-\pi/a<p_\mu\leq\pi/a\}$ and
\begin{equation}
   \Hat{p}_\mu\equiv\frac{2}{a}\sin\left(\frac{ap_\mu}{2}\right),\qquad
   \mathring{p}_\mu\equiv\frac{1}{a}\sin\left(ap_\mu\right).
\label{eq:(4.18)}
\end{equation}
We choose the bare mass parameter~$m_0$ so that the gap equation possesses a
``symmetric solution'' $\Tilde{\sigma}=0$; this corresponds to a massless
fermion because $\Tilde{\sigma}$ provides the fermion mass in the leading order
of the $1/N$ expansion. This requirement leads to
\begin{equation}
   m_0=-0.7698\,\frac{\lambda_0^{\text{LAT}}}{a}.
\label{eq:(4.19)}
\end{equation}
Under this choice, Eq.~\eqref{eq:(4.17)} says that
\begin{equation}
   \Tilde{\sigma}^2=\mathrm{e}^{-1.11861}\mathrm{e}^{-2\pi/\lambda_0^{\text{LAT}}}
   \frac{1}{a^2}.
\label{eq:(4.20)}
\end{equation}
$\Tilde{\sigma}$ has the same physical meaning as $\sigma$
in~Eq.~\eqref{eq:(4.7)}. Thus, by choosing $\lambda_0^{\text{LAT}}$
in~Eq.~\eqref{eq:(4.20)} so that $\Tilde{\sigma}=\sigma$ as a function of~$a$,
and rewriting everything in terms of this renormalized quantity, the dependence
of physical quantities on adopted regularization disappears. In particular, it
can be directly seen that the expression~\eqref{eq:(4.12)} also remains the
same for lattice regularization (with~$\sigma=\Tilde{\sigma}$). In what
follows, we assume that this sort of parameter renormalization is made.

\section{Restoration of  the translation Ward--Takahashi relation}
\label{sec:5}
By using the large $N$ solution in the previous section, we now consider
correlation functions which contain the composite operator~\eqref{eq:(3.16)}.
Then, by studying the small flow-time limit of the correlation functions, we
examine if the energy--momentum tensor defined by our universal formula,
Eq.~\eqref{eq:(3.15)} with~Eq.~\eqref{eq:(3.16)}, fulfills (some particular
cases of) the translation Ward--Takahashi relation, Eq.~\eqref{eq:(3.4)}.

We first note that the fermion propagator in the $1/N$ expansion is,
from~Eqs.~\eqref{eq:(4.1)} and~\eqref{eq:(4.10)},
\begin{equation}
   \left\langle\psi^i(x)\Bar{\psi}^j(y)\right\rangle
   =\delta^{ij}\int_p\mathrm{e}^{ip(x-y)}\,
   \frac{1}{i\Slash{p}+\sigma}+O(1/N)
\label{eq:(5.1)}
\end{equation}
and thus the propagator of the flowed fermion field is given by
\begin{equation}
   \left\langle\chi^i(t,x)\Bar{\chi}^j(s,y)\right\rangle
   =\delta^{ij}\int_p\mathrm{e}^{ip(x-y)}\,
   \frac{\mathrm{e}^{-(t+s)p^2}}{i\Slash{p}+\sigma}+O(1/N)
\label{eq:(5.2)}
\end{equation}
by~Eq.~\eqref{eq:(2.4)}. The
propagator between the flowed and unflowed fermion fields,
$\left\langle\chi^i(t,x)\Bar{\psi}^j(y)\right\rangle$ for example, is given by
simply setting the corresponding flow time zero ($s=0$ in this example)
in~Eq.~\eqref{eq:(5.2)}.

The first correlation function we consider is
\begin{equation}
   \left\langle\partial_\mu\Hat{T}_{\mu\nu}(t,x)\psi^i(y)\Bar{\psi}^i(z)
   \right\rangle.
\label{eq:(5.3)}
\end{equation}

\begin{figure}[ht]
\begin{center}
\includegraphics[scale=0.4,clip]{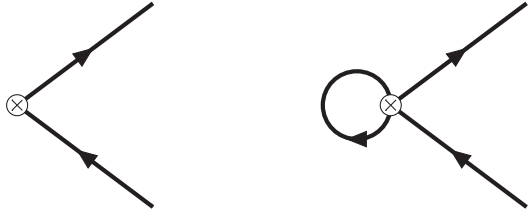}
\caption{Type~I diagrams which contribute to~Eq.~\eqref{eq:(5.3)}. The solid
line is the fermion propagator and the blob denotes the composite
operator~\eqref{eq:(3.16)}.}
\label{fig:1}
\end{center}
\end{figure}

\begin{figure}[ht]
\begin{center}
\includegraphics[scale=0.4,clip]{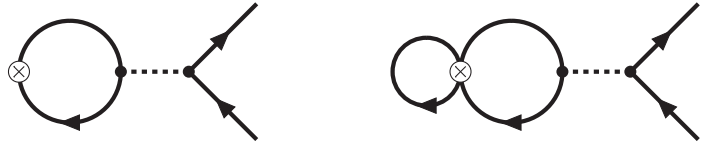}
\caption{Type~II diagrams which contribute to~Eq.~\eqref{eq:(5.3)}. The solid
line is the fermion propagator and the blob denotes the composite
operator~\eqref{eq:(3.16)}. The broken line denotes the propagator of the
auxiliary field, Eq.~\eqref{eq:(4.14)}. The interaction vertex between the
fermion field and the auxiliary field (denoted by the small filled circle) can
be read off from~Eq.~\eqref{eq:(4.1)} with~Eq.~\eqref{eq:(4.10)}.}
\label{fig:2}
\end{center}
\end{figure}

In the leading non-trivial order of the $1/N$ expansion, there are two types of
connected diagrams which contribute to this correlation function; both are
of~$O(N)$. These two types of diagrams are depicted in~Figs.~\ref{fig:1}
and~\ref{fig:2}, respectively.
The contribution of type~I diagrams in~Fig.~\ref{fig:1} is, for small~$t$,
\begin{align}
   &\left\langle\partial_\mu\Hat{T}_{\mu\nu}(t,x)\psi^i(y)\Bar{\psi}^i(z)
   \right\rangle_{\text{I}}
\notag
\\
   &=\int_p\int_q\mathrm{e}^{ip(y-x)}\mathrm{e}^{iq(x-z)}
   N\,\frac{\mathrm{e}^{-tp^2}}{i\Slash{p}+\sigma}
\notag
\\
   &\qquad{}
   \times i(-p+q)_\mu\Biggl\{
   \frac{1}{4}\left[\gamma_\mu i(p+q)_\nu+\gamma_\nu i(p+q)_\mu\right]
   -\frac{1}{2}\delta_{\mu\nu}i(\Slash{p}+\Slash{q})
\notag
\\
   &\qquad\qquad\qquad\qquad{}
   +\frac{\Bar{\lambda}(1/\sqrt{8t})}{2\pi}
   \ln(2\mathrm{e}^\gamma\sigma^2t)
   \left[1+\frac{\Bar{\lambda}(1/\sqrt{8t})}{2\pi}(\ln\pi+1)\right]
   \delta_{\mu\nu}\sigma\Biggr\}
   \frac{\mathrm{e}^{-tq^2}}{i\Slash{q}+\sigma}.
\label{eq:(5.4)}
\end{align}
The second diagram in~Fig.~\ref{fig:1} contains a loop integral arising from
the self-contraction in the last four-fermi term of~Eq.~\eqref{eq:(3.16)}. The
loop integral is finite, however, because of the Gaussian damping factor in the
propagator~\eqref{eq:(5.2)}. We can rewrite the integrand
in~Eq.~\eqref{eq:(5.4)} as
\begin{align}
   &i(-p+q)_\mu\Biggl\{
   \frac{1}{4}\left[\gamma_\mu i(p+q)_\nu+\gamma_\nu i(p+q)_\mu\right]
   -\frac{1}{2}\delta_{\mu\nu}i(\Slash{p}+\Slash{q})
\notag
\\
   &\qquad\qquad\qquad{}
   +\frac{\Bar{\lambda}(1/\sqrt{8t})}{2\pi}
   \ln(2\mathrm{e}^\gamma\sigma^2t)
   \left[1+\frac{\Bar{\lambda}(1/\sqrt{8t})}{2\pi}(\ln\pi+1)\right]
   \delta_{\mu\nu}\sigma\Biggr\}
\notag
\\
   &=(i\Slash{p}+\sigma)
   \left[-iq_\nu
   +\frac{1}{8}[i(-\Slash{p}+\Slash{q}),\gamma_\nu]\right]
   +\left[ip_\nu
   -\frac{1}{8}[i(-\Slash{p}+\Slash{q}),\gamma_\nu]
   \right](i\Slash{q}+\sigma)
\notag
\\
   &\qquad{}
   +\left\{\frac{\Bar{\lambda}(1/\sqrt{8t})}{2\pi}
   \ln(2\mathrm{e}^\gamma\sigma^2t)
   \left[1+\frac{\Bar{\lambda}(1/\sqrt{8t})}{2\pi}(\ln\pi+1)\right]+1\right\}
   i(-p+q)_\nu\sigma.
\label{eq:(5.5)}
\end{align}

On the other hand, from~Eq.~\eqref{eq:(4.9)}, we have
\begin{equation}
   \frac{\Bar{\lambda}(1/\sqrt{8t})}{2\pi}
   =-\frac{1}{\ln(2e^\gamma\sigma^2t/\pi)}
\label{eq:(5.6)}
\end{equation}
and we find the following $t\to0$ limits:
\begin{align}
   &\lim_{t\to0}\frac{\Bar{\lambda}(1/\sqrt{8t})}{2\pi}=0,
\label{eq:(5.7)}
\\
   &\lim_{t\to0}\frac{\Bar{\lambda}(1/\sqrt{8t})}{2\pi}
   \ln(2\mathrm{e}^\gamma\sigma^2t)=-1,
\label{eq:(5.8)}
\\
   &\lim_{t\to0}
   \ln(2\mathrm{e}^\gamma\sigma^2t)
   \left[1+\frac{\Bar{\lambda}(1/\sqrt{8t})}{2\pi}
   \ln(2\mathrm{e}^\gamma\sigma^2t)\right]
   =-\ln\pi,
\label{eq:(5.9)}
\\
   &\lim_{t\to0}\left[\frac{\Bar{\lambda}(1/\sqrt{8t})}{2\pi}
   \ln(2\mathrm{e}^\gamma\sigma^2t)\right]^2=1.
\label{eq:(5.10)}
\end{align}
From these, we see that the last line of Eq.~\eqref{eq:(5.5)} vanishes
for~$t\to0$. Then when Eq.~\eqref{eq:(5.5)} is substituted
in~Eq.~\eqref{eq:(5.4)}, the factor~$(i\Slash{p}+\sigma)$
in~Eq.~\eqref{eq:(5.5)} cancels the external propagator $1/(i\Slash{p}+\sigma)$
and then the integration over~$p$ produces the delta function $\delta^2(x-y)$
for~$t\to0$. The situation is similar for the factor~$(i\Slash{q}+\sigma)$
in~\eqref{eq:(5.5)}. In this way, we have
\begin{align}
   &\lim_{t\to0}
   \left\langle\partial_\mu\Hat{T}_{\mu\nu}(t,x)\psi^i(y)\Bar{\psi}^i(z)
   \right\rangle_{\text{I}}
\notag
\\
   &=-\delta^2(x-y)
   \left\langle\partial_\nu\psi^i(y)\Bar{\psi}^i(z)\right\rangle
   -\delta^2(x-z)
   \left\langle\psi^i(y)\partial_\nu\Bar{\psi}^i(z)\right\rangle
\notag
\\
   &\qquad{}
   +\partial_\mu
   \left[\delta^2(x-y)\frac{1}{8}[\gamma_\mu,\gamma_\nu]
   \left\langle\psi^i(y)\Bar{\psi}^i(z)\right\rangle
   -\delta^2(x-z)
   \left\langle\psi^i(y)\Bar{\psi}^i(z)\right\rangle
   \frac{1}{8}[\gamma_\mu,\gamma_\nu]
   \right]+O(N^0).
\label{eq:(5.11)}
\end{align}
This is precisely the expected form of the Ward--Takahashi relation associated
with the translational invariance. In fact, by considering the integration
over the position~$x$ over the region that contains the points $y$ and~$z$, we
observe that Eq.~\eqref{eq:(3.4)}
with~$\mathcal{O}_{\text{int}}=\psi^i(y)\Bar{\psi}^i(z)$
(and~$\mathcal{O}_{\text{ext}}=1$) holds.\footnote{The energy--momentum tensor
always possesses the ambiguity that results in the total divergence in the
(unintegrated) Ward--Takahashi relation associated with the translational
invariance. The second line of~Eq.~\eqref{eq:(5.11)}, which corresponds to the
Lorentz rotation generated by the anti-symmetric part of the canonical
energy--momentum tensor, being the total divergence, does not contribute to the
integrated form of the translation Ward--Takahashi relation,
Eq.~\eqref{eq:(3.4)}.}

\begin{figure}[ht]
\begin{center}
\includegraphics[scale=0.4,clip]{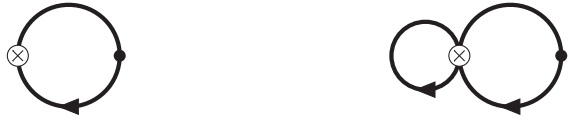}
\caption{
The left-hand side parts of the type~II diagrams in~Fig.~\ref{fig:2}.}
\label{fig:3}
\end{center}
\end{figure}

Since the correct Ward--Takahashi relation is already saturated by type~I
diagrams, Eq.~\eqref{eq:(5.11)}, the type~II diagrams in~Fig.~\ref{fig:2}
should not contribute to the translation Ward--Takahashi identity. To see this,
and for a later use, it is useful to compute first the left-hand side parts of
the type~II diagrams depicted in~Fig.~\ref{fig:3}.
An explicit computation of the diagrams in~Fig.~\ref{fig:3} yields
\begin{align}
   &\int_r\mathrm{e}^{ir(x-y)}\frac{N}{2\pi}\sigma
   \Biggl(
   \left(\delta_{\mu\nu}-\frac{r_\mu r_\nu}{r^2}\right)B(r^2,\sigma^2)
   -\delta_{\mu\nu}+2\frac{r_\mu r_\nu}{r^2}
\notag
\\
   &{}
   -\delta_{\mu\nu}
   \left\{
   1+\frac{\Bar{\lambda}(1/\sqrt{8t})}{2\pi}\ln(2\mathrm{e}^\gamma\sigma^2t)
   \left[1+\frac{\Bar{\lambda}(1/\sqrt{8t})}{2\pi}(\ln\pi+1)\right]\right\}
   \left[\ln(2\mathrm{e}^\gamma\sigma^2t)+B(r^2,\sigma^2)\right]
   \Biggr).
\label{eq:(5.12)}
\end{align}
Using Eqs.~\eqref{eq:(5.7)}--\eqref{eq:(5.10)}, we then have
\begin{equation}
   \lim_{t\to0}\text{Eq.~\eqref{eq:(5.12)}}
   =\int_r\mathrm{e}^{ir(x-y)}\frac{N}{2\pi}\sigma
   \left(\delta_{\mu\nu}-\frac{r_\mu r_\nu}{r^2}\right)
   \left[B(r^2,\sigma^2)-2\right].
\label{eq:(5.13)}
\end{equation}
From this, for the type~II diagrams in~Fig.~\ref{fig:2}, 
\begin{align}
   &\lim_{t\to0}\left\langle\Hat{T}_{\mu\nu}(t,x)\psi^i(y)\Bar{\psi}^i(z)
   \right\rangle_{\text{II}}
\notag
\\
   &=\int_p\int_q\mathrm{e}^{ip(y-x)}\mathrm{e}^{iq(x-z)}
   N\sigma
   \left(\delta_{\mu\nu}-\frac{r_\mu r_\nu}{r^2}\right)
   \left[1-2B(r^2,\sigma^2)^{-1}\right]
   \frac{1}{i\Slash{p}+\sigma}\frac{1}{i\Slash{q}+\sigma},
\label{eq:(5.14)}
\end{align}
where
\begin{equation}
   r\equiv-p+q,
\label{eq:(5.15)}
\end{equation}
and we have the desired result
\begin{equation}
   \lim_{t\to0}
   \left\langle\partial_\mu\Hat{T}_{\mu\nu}(t,x)\psi^i(y)\Bar{\psi}^i(z)
   \right\rangle_{\text{II}}=0.
\label{eq:(5.16)}
\end{equation}

Thus, in the leading non-trivial order of the $1/N$ expansion, we have
confirmed that the universal formula, Eq.~\eqref{eq:(3.15)}
with~Eq.~\eqref{eq:(3.16)}, reproduces the translation Ward--Takahashi
relation~\eqref{eq:(3.4)} for the product of elementary fields,
$\mathcal{O}_{\text{int}}=\psi^i(y)\Bar{\psi}^i(z)$
(and~$\mathcal{O}_{\text{ext}}=1$). This shows that the universal formula
reproduces the correct normalization and the conservation law for the
energy--momentum tensor, at least in the correlation function with elementary
fields.

\begin{figure}[ht]
\begin{center}
\includegraphics[scale=0.4,clip]{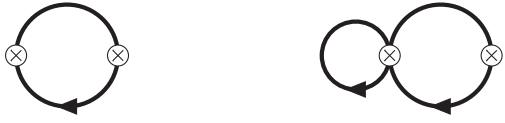}
\caption{Type~I diagrams which contribute to~Eq.~\eqref{eq:(5.18)}. In each
diagram, the left blob denotes the composite operator~\eqref{eq:(3.16)} and the
right blob denotes the scalar density operator in~Eq.~\eqref{eq:(5.17)}.}
\label{fig:4}
\end{center}
\end{figure}

\begin{figure}[ht]
\begin{center}
\includegraphics[scale=0.4,clip]{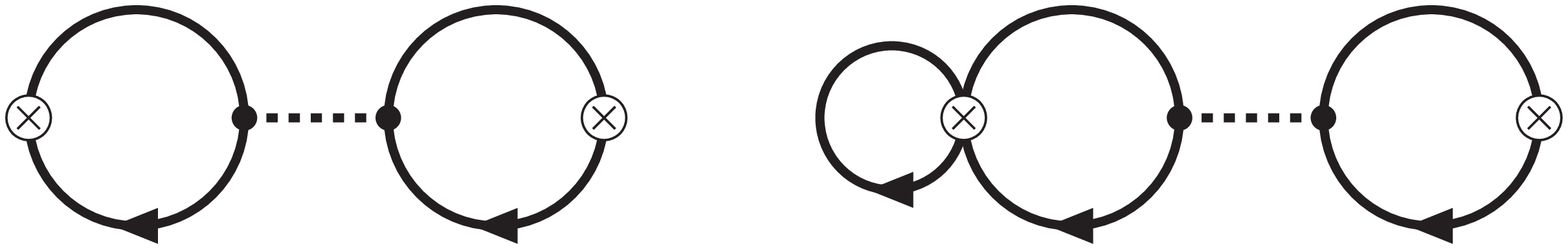}
\caption{Type~II diagrams which contribute to~Eq.~\eqref{eq:(5.18)}. In each
diagram, the left blob denotes the composite operator~\eqref{eq:(3.16)} and the
right blob denotes the scalar density operator in~Eq.~\eqref{eq:(5.17)}.}
\label{fig:5}
\end{center}
\end{figure}

The above computation in fact demonstrates that Eq.~\eqref{eq:(3.4)} is also
reproduced for the scalar density operator, that is,
\begin{equation}
   \mathcal{O}_{\text{int}}=Z_S\Bar{\psi}^i(y)\psi^i(y),\qquad
   \mathcal{O}_{\text{ext}}=1,
\label{eq:(5.17)}
\end{equation}
where $Z_S$ is an appropriate renormalization factor for the scalar density. In
the leading non-trivial order of the $1/N$ expansion, there are two types of
diagrams which contribute to
\begin{equation}
   \left\langle\partial_\mu\Hat{T}_{\mu\nu}(t,x)\Bar{\psi}^i(y)\psi^i(y)
   \right\rangle,
\label{eq:(5.18)}
\end{equation}
as depicted in~Figs.~\ref{fig:4} and~\ref{fig:5}.

For type~I diagrams, the computation is identical to that
for~Eqs.~\eqref{eq:(5.12)} and~\eqref{eq:(5.13)}. As is clear
from~Eq.~\eqref{eq:(5.13)}, we have
$\lim_{t\to0}\langle\partial_\mu\Hat{T}_{\mu\nu}(t,x)\Bar{\psi}^i(y)\psi^i(y)\rangle_{\text{I}}=0$. For type~II diagrams also, we do not need a new calculation
because Eqs.~\eqref{eq:(5.12)} and~\eqref{eq:(5.13)} (which correspond
to~Fig.~\ref{fig:3}) give the parts of the diagrams in~Fig.~\ref{fig:5}. Thus,
again from~Eq.~\eqref{eq:(5.13)}, we have
$\lim_{t\to0}\langle\partial_\mu\Hat{T}_{\mu\nu}(t,x)\Bar{\psi}^i(y)\psi^i(y)\rangle_{\text{II}}=0$. These reproduce Eq.~\eqref{eq:(3.4)}
with~Eq.~\eqref{eq:(5.17)} because
$\langle\partial_\nu[\Bar{\psi}^i(y)\psi^i(y)]\rangle=0$ by the translational
invariance.

\begin{figure}[ht]
\begin{center}
\includegraphics[scale=0.4,clip]{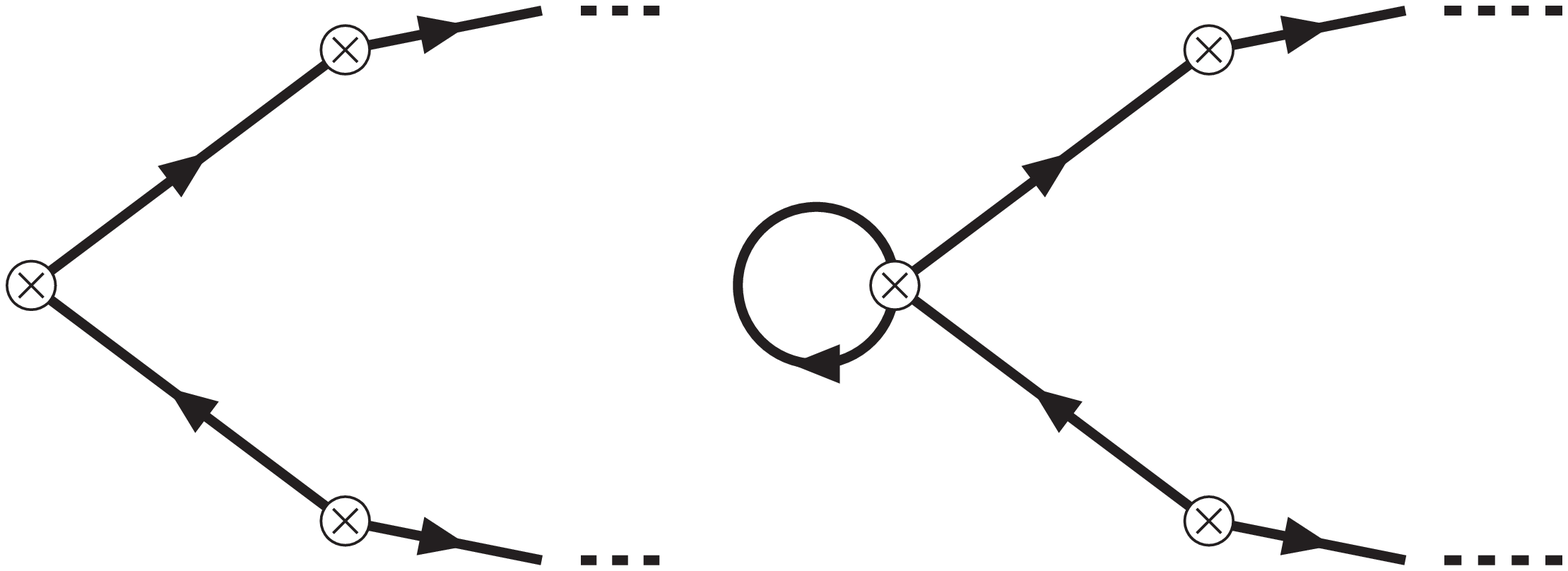}
\caption{Type~I diagrams which contribute to~Eq.~\eqref{eq:(5.19)}. In each
diagram, the leftmost blob denotes the composite operator~\eqref{eq:(3.16)} and
other blobs denote the fermion composite operators contained
in~$\mathcal{O}_{\text{ext}}$ in~Eq.~\eqref{eq:(5.19)}.}
\label{fig:6}
\end{center}
\end{figure}

\begin{figure}[ht]
\begin{center}
\includegraphics[scale=0.4,clip]{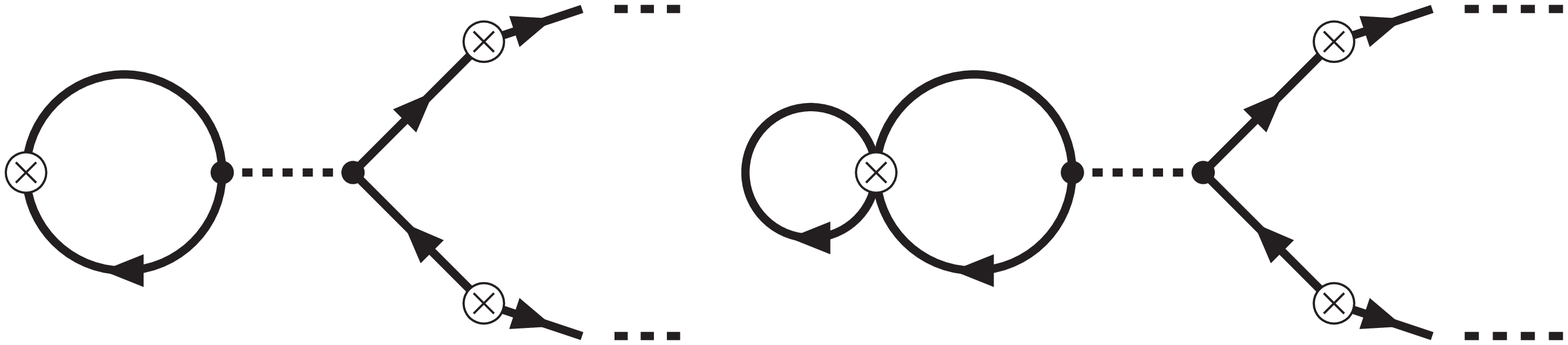}
\caption{Type~II diagrams which contribute to~Eq.~\eqref{eq:(5.19)}. In each
diagram, the leftmost blob denotes the composite operator~\eqref{eq:(3.16)} and
other blobs denote the fermion composite operators contained
in~$\mathcal{O}_{\text{ext}}$ in~Eq.~\eqref{eq:(5.19)}.}
\label{fig:7}
\end{center}
\end{figure}

We can further argue that Eq.~\eqref{eq:(3.4)} is reproduced when
$\mathcal{O}_{\text{ext}}$ is a collection of renormalized composite operators
of the fermion field and~$\mathcal{O}_{\text{int}}=1$. That is, for this
situation, we can argue that
\begin{equation}
   \lim_{t\to0}
   \left\langle\mathcal{O}_{\text{ext}}\int_{\mathcal{D}}\mathrm{d}^Dx\,
   \partial_\mu\Hat{T}_{\mu\nu}(t,x)\right\rangle=0.
\label{eq:(5.19)}
\end{equation}
This shows that the conservation law of the energy--momentum tensor is
reproduced in the correlation function with generic composite operators. The
argument is simple: There exist two types of diagrams which contribute
to~Eq.~\eqref{eq:(5.19)}. For type~I diagrams in~Fig.~\ref{fig:6}, we can use
the identity~\eqref{eq:(5.5)} for fermion lines starting from the vertex of the
composite operator~\eqref{eq:(3.16)}. Then, as~Eq.~\eqref{eq:(5.11)}, we have
\begin{align}
   &\lim_{t\to0}
   \left\langle
   \psi^i(y)\Bar{\psi}^i(z)
   \dotsm
   \partial_\mu\Hat{T}_{\mu\nu}(t,x)
   \right\rangle_{\text{I}}
\notag
\\
   &=-\delta^2(x-y)
   \left\langle\partial_\nu\psi^i(y)\Bar{\psi}^i(z)\dotsm\right\rangle
   -\delta^2(x-z)
   \left\langle\psi^i(y)\partial_\nu\Bar{\psi}^i(z)\dotsm\right\rangle
   -\dotsb
\notag
\\
   &\qquad{}
   +\left[
   \text{$\delta^2(x-y)$, $\delta^2(x-z)$, \dots, inside the total divergence
   in~$x$}
   \right].
\label{eq:(5.20)}
\end{align}
Then for $x\neq y$, $x\neq z$, \dots, the right-hand side vanishes. For the
type~II diagrams in~Fig.~\ref{fig:7}, from~Eq.~\eqref{eq:(5.13)}, we simply
have $\lim_{t\to0}\langle\mathcal{O}_{\text{ext}}
\partial_\mu\Hat{T}_{\mu\nu}(t,x)\rangle_{\text{II}}=0$. These imply
Eq.~\eqref{eq:(5.19)}.

\section{Expectation value at finite temperature}
\label{sec:6}
The Ward--Takahashi relation~\eqref{eq:(5.11)} shows that our universal formula
for the energy--momentum tensor gives rise to the correct normalization at
least within the correlation function with elementary fields. To give a further
support on the correct normalization, in this section we compute the
expectation value of the composite operator~\eqref{eq:(3.16)} at finite
temperature and compare it with thermodynamic quantities directly obtained in
the conventional $1/N$ expansion. A similar analysis for the two-dimensional
$O(N)$ non-linear sigma model has been carried out
in~Ref.~\cite{Makino:2014cxa}.

At finite temperature with inverse temperature~$\beta$, the propagator is given
by
\begin{equation}
   \left\langle\chi^i(t,x)\Bar{\chi}^j(s,y)\right\rangle_\beta
   =\delta^{ij}\frac{1}{\beta}\sum_{n=-\infty}^\infty
   \int\frac{\mathrm{d}p_1}{2\pi}\,\mathrm{e}^{ip(x-y)}
   \frac{\mathrm{e}^{-(t+s)(\omega_n^2+p_1^2)}}
   {i\gamma_0\omega_n+i\gamma_1p_1+\sigma_\beta}+O(1/N),
\label{eq:(6.1)}
\end{equation}
where $\omega_n$ is the Matsubara frequency
\begin{equation}
   \omega_n\equiv\frac{2\pi n}{\beta},
\label{eq:(6.2)}
\end{equation}
and $\sigma_\beta$ is the large $N$ saddle point at finite temperature;
$\sigma_\beta$~is given by a finite-temperature counterpart of the gap
equation~\eqref{eq:(4.3)}:
\begin{equation}
   \frac{1}{\lambda_0}
   =\tr\frac{1}{\beta}\sum_{n=-\infty}^\infty
   \int\frac{\mathrm{d}p_1}{2\pi}\,
   \frac{1}{i\gamma_0\omega_n+i\gamma_1p_1+\sigma_\beta}.
\label{eq:(6.3)}
\end{equation}

The required computation is almost the same as that
in~Ref.~\cite{Makino:2014cxa}, because of the similarity of expressions. Using
Eq.~\eqref{eq:(3.15)} with~Eq.~\eqref{eq:(3.16)}, for the energy
density~$\varepsilon$ we have
\begin{align}
   \varepsilon&=-\left\langle\left\{T_{00}\right\}_R(x)\right\rangle_\beta
\notag\\
   &=-\frac{N}{4\pi}(\sigma_\beta^2-\sigma^2)
   -\frac{N}{\pi}\sigma_\beta^2\sum_{n=1}^\infty K_2(\beta\sigma_\beta n)
   +O(N^0)
\label{eq:(6.4)}
\end{align}
and, for the pressure~$P$,
\begin{align}
   P&=\left\langle\left\{T_{11}\right\}_R(x)\right\rangle_\beta
\notag\\
   &=\frac{N}{4\pi}(\sigma_\beta^2-\sigma^2)
   -\frac{N}{\pi}\sigma_\beta^2\sum_{n=1}^\infty K_2(\beta\sigma_\beta n)
   +O(N^0).
\label{eq:(6.5)}
\end{align}
These are the results of the universal formula.

On the other hand, the free energy density of the present system is given by
\begin{equation}
   f(\beta)=\frac{N}{2\lambda_0}\beta\sigma_\beta^2
   -N\frac{1}{\beta}\sum_{n=-\infty}^\infty
   \int\frac{\mathrm{d}p_1}{2\pi}\,
   \ln\left(\omega_n^2+p_1^2+\sigma_\beta^2\right),
\label{eq:(6.6)}
\end{equation}
and the energy-density and the pressure are given by
$\varepsilon=\partial f(\beta)/\partial\beta$ and~$P=-f(\beta)/\beta$,
respectively. From comparison of~Eq.~\eqref{eq:(6.6)} with~Eq.~(A1)
of~Ref.~\cite{Makino:2014cxa}, we see that these quantities can be obtained by
making the substitutions $f(\beta)\to-f(\beta)$, $N\to2N$,
$\lambda_0\to2\lambda_0$, $\sigma_\beta\to\sigma_\beta^2$,
and~$\sigma\to\sigma^2$ in Eqs.~(A11) and~(A9) of~Ref.~\cite{Makino:2014cxa}.
We then observe a complete agreement with Eqs.~\eqref{eq:(6.4)}
and~\eqref{eq:(6.5)}. This result again supports the validity of our universal
formula for the energy--momentum tensor.

\section{Conclusion}
\label{sec:7}
The flow in quantum field theory and the small flow-time expansion can give
rise to a regularization-independent expression for composite operators. In
this paper, we examined the validity of a universal formula for the
energy--momentum tensor by using the Gross--Neveu model and the
non-perturbative $1/N$ expansion. In the leading non-trivial order in the $1/N$
expansion, we have observed that (some particular cases of) the Ward--Takahashi
relation associated with the translational invariance is correctly reproduced
by the universal formula even with the non-perturbative mass gap. This is
interesting because the construction of the universal formula itself requires
only (one-loop) perturbation theory. We have also observed that the formula
reproduces thermodynamic quantities correctly. These observations support the
validity of a similar construction of the energy--momentum tensor via the
gradient/Wilson flow in lattice gauge theory.

\section*{Acknowledgment}

The work of H.~S. is supported in part by Grant-in-Aid for Scientific
Research~23540330.

\section*{Note added}
I learned that a strikingly analogous idea to define the energy--momentum
tensor to ours has been considered on the basis of the operator product
expansion (instead of the small flow-time expansion)
in~Ref.~\cite{Holland:2013xya}. I would like to thank Jan Holland for the
information.


\end{document}